\documentclass[11pt]{article}
\pdfoutput=1

\usepackage[a4paper,margin=1in]{geometry}
\usepackage{mathtools}
\usepackage{graphicx}
\usepackage{url} 
\usepackage{float}
\usepackage{setspace, bm} 
\usepackage{appendix}
\usepackage{enumerate, enumitem}
\usepackage{subcaption} 
\usepackage{threeparttable}

\usepackage{multirow}
\usepackage{verbatim}
\usepackage{amssymb}
\usepackage[table]{xcolor}
\usepackage{booktabs}
\usepackage{changepage}
\usepackage[T1]{fontenc}

\usepackage{rotating} 

\usepackage{mathtools}
\usepackage{graphicx}
\usepackage{url} 
\usepackage{float}
\usepackage{setspace, bm} 

\usepackage{enumerate, enumitem}

\usepackage{rotating} 

\usepackage{algorithm}
\usepackage{algpseudocode}

\def\bSig\mathbf{\Sigma}

\usepackage{amssymb, amsmath}
\newtheorem{theorem}{Theorem}
\newtheorem{corollary}{Corollary}
\newtheorem{assumption}{Assumption}
\newtheorem{lemma}{Lemma}
\newtheorem{remark}{Remark}%

\usepackage{siunitx}
\PassOptionsToPackage{hyphens}{url}\usepackage[hidelinks]{hyperref}
\usepackage{cleveref}
\usepackage[utf8]{inputenc}
\usepackage[right]{lineno}
\usepackage{csquotes}
\usepackage{booktabs}
\usepackage{longtable}
\usepackage{adjustbox}
\usepackage{array}
\usepackage{titlesec}
\usepackage{authblk}
\usepackage{xcolor} 

\usepackage{listings}
\usepackage{multirow}

\usepackage{listings}

\usepackage{xcolor}
\definecolor{codegreen}{rgb}{0,0.6,0}
\definecolor{codegray}{rgb}{0.5,0.5,0.5}
\definecolor{codepurple}{rgb}{0.58,0,0.82}
\definecolor{backcolour}{rgb}{0.95,0.95,0.92}

\usepackage{tcolorbox}

\usepackage{longtable}
\usepackage{wrapfig}
\usepackage{makecell}

\usepackage{amsmath}

\usepackage[utf8]{inputenc} 
\usepackage[T1]{fontenc}    
\usepackage{hyperref}       
\usepackage{url}            
\usepackage{booktabs}       
\usepackage{amsfonts}       
\usepackage{nicefrac}       
\usepackage{microtype}      
\usepackage{subcaption} 
\usepackage[symbol]{footmisc}
\usepackage{tcolorbox}
\tcbuselibrary{listings,breakable}
\usepackage{verbatimbox}
\usepackage{minted}
\usepackage{mdframed}
\usepackage{todonotes}

\usepackage{algorithm}
\usepackage{algpseudocode}
\usepackage{graphicx}

\usepackage{amssymb}

\def\bA{\mathbf{A}}

\def\bE{\mathbf{E}}

\def\bN{\mathbf{N}}

\def\bP{\mathbf{P}}

\def\bV{\mathbf{V}}

\def\bbP{\mathbb{P}}

\def\bbR{\mathbb{R}}

\def\cO{\mathcal{O}}

\def\cR{\mathcal{R}}

\def\cX{\mathcal{X}}
\def\cY{\mathcal{Y}}

\def\Ex{{\bE}}

\def\eg{{\it e.g.}}

\def\debias{\mathrm{debias}}

\def\var{\mathrm{var}}

\usepackage{nicefrac}       
\usepackage{microtype}      
\usepackage{graphicx}
\usepackage{natbib}
\usepackage{doi}

\usepackage[english]{babel}

\title{Robust inference for risk heterogeneity under group imbalance}
\author{Mengqi Xu\thanks{m332xu@uwaterloo.ca}}
\author{Subha Maity\thanks{subha.maity@uwaterloo.ca}}
\author{Joel Dubin\thanks{jdubin@uwaterloo.ca}}

\affil{Department of Statistics and Actuarial Science \\
University of Waterloo}

\date{}

\def\bSig\mathbf{\Sigma}

\begin{document}
{\singlespacing
\maketitle}
\vspace*{-3em}

\begin{abstract}

Population-level heterogeneity is ubiquitous in biomedical data, where differences across demographic or clinical subgroups can substantially alter risk patterns. For example, in intensive care unit (ICU) studies, the mortality risk associated with specific admission diagnoses can vary across ethnic groups. Existing approaches for detecting risk heterogeneity are often sensitive to baseline model misspecification and regularization bias, both of which are aggravated when one subgroup is severely under-represented, as is common in practice. In this paper, we propose a robust framework for inferring risk heterogeneity between two populations using Neyman orthogonality, which yields estimators that are locally insensitive to nuisance parameter estimation error. The proposed estimator is consistent and asymptotically normal, and simulation studies demonstrate that in finite samples our method substantially reduces bias and improves inferential stability compared with standard likelihood-based approaches. In an application to the eICU Collaborative Research Database, our method reveals clinically meaningful ethnicity-specific heterogeneity in admission diagnoses for in-hospital mortality that penalized and unpenalized likelihood-based methods fail to detect.


\end{abstract}

\section{Introduction}\label{s:intro}


In many healthcare applications, demographic, environmental, and contextual factors drive significant disparities in patient outcomes across different subgroups. These observed disparities, however, are highly sensitive to the scale of analysis and can fluctuate between the aggregate population level and specific sub-cohorts. For example, the eICU Collaborative Research Database \citep{pollard2018eicu}, a large publicly-available database of intensive care unit (ICU) patients in the US, indicates that the overall in-hospital mortality prevalence is slightly higher for ``Caucasian'' patients (9.77\%) than for ``Asian'' patients (9.39\%). Conversely, among patients whose primary admission diagnosis is cardiac arrest, this disparity reverses, with Asian patients facing a much higher mortality prevalence (40.74\%) than Caucasian patients (21.29\%) (Table \ref{ch3tab:AsianCA}). 
Covariates may also further influence such disparities between ethnicites across different diseases. Therefore, understanding and accounting for outcome heterogeneity at a granular, covariate-adjusted level is a fundamental prerequisite for building clinical models that are generalizable, interpretable, and precise.

\begin{table}[h]
\centering
\caption{In-hospital mortality of Caucasian and Asian patients in the eICU dataset, overall and within the cardiac arrest subgroup. Prevalence is the percentage of expired patients within each ethnicity and subgroup.}
\label{ch3tab:AsianCA}
\begin{tabular}{lrrrr}
\toprule
& \multicolumn{2}{c}{\textbf{Overall}} & \multicolumn{2}{c}{\textbf{Cardiac Arrest}} \\
\cmidrule(lr){2-3} \cmidrule(lr){4-5}
\textbf{Status} & \textbf{Caucasian} & \textbf{Asian} & \textbf{Caucasian} & \textbf{Asian} \\
\midrule
Alive    & 33427 & 473 & 2336 & 16 \\
Expired  & 3616  & 49  & 632  & 11 \\
\midrule
Total    & 37043 & 522 & 2968 & 27 \\
\midrule
Prevalence (\%) & 9.77 & 9.39 & 21.29 & 40.74 \\
\bottomrule
\end{tabular}
\end{table}

Prior work has documented substantial mortality differences across ethnicities \citep{mcgowan2022racial}, heterogeneous treatment effects in subgroup analyses \citep{rekkas2020predictive}, and potential inequities in intensive care unit (ICU) interventions among diverse sociodemographic populations \citep{kotfis2024equity}. 
Simple test statistics such as z-scores, t-tests, or chi-squared tests \citep{marusteri2010comparing} provide group comparisons but fail to adjust for multiple factors. While multivariable frameworks like ANCOVA \citep{schneider2015cautionary} or baseline category logit models \citep{kwak2002multinomial, agresti2002categorical} can adjust for covariates, they struggle with a large number of predictors and when one of the subgroups is under-represented \citep{albert1984existence}.
Moreover, they do not provide a flexible framework to account for any non-linear associations that may exist between the outcome and the covariates.

Compared to prior work, we aim to develop an inferential framework to understand the covariate-adjusted heterogeneity across subgroups that are amenable to a large number of covariates, with the potential underrepresentation of some subgroups, and can flexibly account for non-linear associations. For our framework, we consider the following example model: 
\begin{equation}\label{ch3eq:la}
    \mathrm{logit}\big\{\bP(Y = 1|\boldsymbol{X},\text{Asian})\big\} -\mathrm{logit}\big \{\bP(Y = 1| \boldsymbol{X},\text{Caucasian})\big \} = \alpha^T \boldsymbol{X}\,,
\end{equation} where 
$Y$ represents binary in-hospital mortality and $\boldsymbol X$ represents a covariate vector. Thus, in \eqref{ch3eq:la}, heterogeneity is formulated as a linear function of $\boldsymbol X$.
Such a model was considered by \citet{maity2024linear} to predict mortality for an ethnically under-represented subgroup by leveraging information from a well-represented subgroup; hence, it is suited for dealing with subgroup underrepresentation. 
Although we posit the difference in the logits as a linear function of $\boldsymbol{X}$, the logits individually may be non-linear functions, thus allowing for non-linear associations between covariates and outcomes. Moreover, the model provides an interpretable way to  compare subgroup-specific associations after covariate adjustment. For example, if the $j$th covariate is an indicator for cardiac arrest, then $\alpha_j$ represents the marginal difference in log odds of mortality between Asians and Caucasians when an individual is admitted due to cardiac arrest, after adjusting for other covariates. To statistically verify whether this difference is significant, we want to perform inference on $\alpha_j$.





Conducting reliable heterogeneity inference in modern healthcare studies is often hampered by two challenges. First, one of the subgroups may be severely under-represented; \eg\ in the eICU data, ``Asian'' ethnicity has only 522 samples, of whom only 49 died. This underrepresentation is further exacerbated by the large number of covariates.
Second, existing methods for the model in Eq. \eqref{ch3eq:la}, such as \citet{maity2024linear}, typically rely on regularized likelihood-based estimation, which yields estimators of $\alpha_j$ that can be sensitive to baseline model mis-specification (in our case, the Caucasian model) and regularization bias under a high-dimensional setting. This is unsuitable for inferential purposes. 
Simply removing the regularization is not a remedy: the sensitivity to the baseline model remains, and with few events relative to the number of covariates, the unpenalized estimates become unstable or fail to exist \citep{albert1984existence}.


To address these limitations, we draw inspiration from Neyman orthogonalization \citep{chernozhukov2018double}, which allows for the construction of estimators that are robust to estimation errors in ``nuisance'' components of the model, such as the baseline model for the majority group $\mathrm{logit}\big \{\bP(Y = 1| \boldsymbol{X},\text{Caucasian})\big \}$ and other auxiliary parameters. Neyman orthogonalization has been widely applied in causal inference \citep{chernozhukov2018double,oprescu2019orthogonal}
and provides a successful tool for mitigating biases introduced by nuisance estimators. By leveraging this approach, we propose a flexible and robust inference framework that ensures reliable inference on the heterogeneities, even when the estimates of the baseline model and nuisance parameters are biased. 


The proposed framework offers several key contributions: (i) it provides an inference tool that can handle under-represented subgroups; (ii) it accommodates a large number of covariates; (iii) it is highly flexible with respect to the choice of baseline model algorithm; and, finally, (iv) it extends beyond binary outcomes such as in-hospital mortality to a broad range of outcome types. 
The remainder of this paper is organized as follows. In Section \ref{s:setup}, we introduce the problem setup and our method. Specifically, in Section \ref{s:method}, we propose the novel method by introducing a Neyman orthogonal score function in our setting, and in
Section \ref{s:inference}, we establish the theoretical property of the proposed estimator.
Then, in Section \ref{s:simulation}, we present simulation results that illustrate the debiasing advantage of our method over standard linear adjustment approaches. In Section \ref{s:application}, we apply the proposed framework to the eICU Collaborative Research Database \citep{pollard2018eicu} to infer ethnic heterogeneities in ICU admission diagnoses for in-hospital mortality prediction. Finally, in Section \ref{s:discussion}, we conclude the paper with a discussion.

\section{Methodology}\label{s:setup}
\subsection{Problem setup}

Here, we formalize the problem setup. We assume that the dataset  $\{(\boldsymbol{X}_i, Y_i, T_i)\}_{i = 1}^N$ is a random sample drawn from a probability distribution $\bbP$, where $\boldsymbol{X}_i \in \cX \subseteq \bbR^d$ and $Y_i\in \cY$ are the predictors and outcomes of interest, and $\cX$ and $\cY$ denote the spaces of the predictors and outcomes, respectively.  Moreover, $T_i \in \{0, 1\}$ indicates the subpopulation label; where the well-represented group is denoted as $T_i = 1$ and the under-represented group is denoted as $T_i = 0$. Consequently, $\bbP(T_i = 1) \ge \nicefrac{1}{2}$, often closer to 1 in many applications. 

Recall that our primary goal is to perform statistical inference for the parameters defined in Eq. \eqref{ch3eq:la}. Throughout this paper, we are primarily interested in a binary outcome (in-hospital mortality of ICU patients), and these parameters are defined using the logistic link function, as in \eqref{ch3eq:la}. However, for a broader exposition, we shall describe our methods with a general, prespecified link function $\gamma$. To this point, we denote the link-transformed regression function for the majority subpopulation, which is $\gamma(\Ex_{\bbP}[Y_i\mid \boldsymbol{X}_i, T_i = 1])$, as $\xi(\boldsymbol{X}_i)$. Furthermore, for $y \in \cY$ and $a \in \bbR$, consider the loss function $\ell(a, y) = -ya + \Psi(a)$ such that $\Psi' (a) = \gamma^{-1}(a)$, which is frequently used in generalized linear models with the natural link functions. Henceforth, we shall denote $\gamma^{-1}$ as $\sigma$. 
As some examples, we may consider $\ell(y, a) = - ya + \nicefrac{a^2}{2}$ for linear regression with the identity link function and $\ell(a, y) = -ya  + \log(1 + e^a)$ for logistic regression with the logistic link function. We shall use this latter example in our simulation (Section \ref{s:simulation}) and in the data application (Section \ref{s:application}). 

As described in Eq. \eqref{ch3eq:la}, for our comparative study, we shall quantify the difference $\gamma(\Ex_{\bbP}[Y_i\mid \boldsymbol{X}_i, T_i = 0]) - \gamma(\Ex_{\bbP}[Y_i\mid \boldsymbol{X}_i, T_i = 1])$ through $\alpha^\top \boldsymbol{X}_i$, a linear function of $\boldsymbol{X}_i$. Thus, we consider the working model 
\begin{equation}\label{eq:working-model}
    \gamma(\Ex_{\bbP}[Y_i\mid \boldsymbol{X}_i, T_i = 0]) = \gamma(\Ex_{\bbP}[Y_i\mid \boldsymbol{X}_i, T_i = 1]) + \alpha^\top \boldsymbol{X}_i  = \xi(\boldsymbol{X}_i) + \alpha^\top \boldsymbol{X}_i\,,
\end{equation}
 and define the population-level values of $\xi(\cdot)$ and $\alpha$ as
\begin{equation} \label{eq:population-optimization}
\begin{aligned}
    \alpha^{\star}, \xi^{\star}(\cdot) = \underset{\alpha \in \bbR^d, \xi \in \Xi}{\arg\min}\; \Ex_{\bbP} \big[\psi (\boldsymbol{X}_i, Y_i , T_i, \alpha, \xi)\big]\, ,
\end{aligned}
\end{equation}  
where \(\psi (\boldsymbol{Z}_i, \alpha, \xi) := T_i\ell (\xi(\boldsymbol{X}_i), Y_i) + (1 - T_i) \ell (\xi(\boldsymbol{X}_i) +\alpha^\top \boldsymbol{X}_i, Y_i)\, , \,\)
$\Xi$ is an appropriate function class, and $\boldsymbol{Z}_i := (\boldsymbol{X}_i, Y_i, T_i)$.

We focus on conducting statistical inference for a subset of parameters in $\alpha$: given a set $S \subseteq \{1, \dots, d\}$, our goal is to perform statistical inference on $\beta = \{\alpha_j : j \in S\}$. As a particular instance, taking $S = \{1, \dots, d\}$ yields an inference for the entire vector $\alpha$. Moreover, when $\boldsymbol{X}_i$ is high-dimensional, practitioners are typically interested in a prespecified low-dimensional subset $\beta$, rather than the full high-dimensional vector $\alpha$. Thus, our parameter of interest (vector) is $\beta$, and we shall refer to $\xi(\cdot)$ and $\delta = \{\alpha_j : j \in S^\complement\}$ as nuisance parameters. Recall that $\alpha^\star$ and $\xi^\star$ are the population-level values for $\alpha$ and $\xi$, respectively, and similarly, we define the population-level values for $\beta$ and $\delta$ as $\beta^\star = \{\alpha_j^\star : j \in S\}$ and $\delta^\star = \{\alpha_j^\star : j \in S^\complement\}$, respectively. We also define $\boldsymbol{U}_i = \{\boldsymbol{X}_{i, j}: j \in S\}$ and $\boldsymbol{V}_i = \{\boldsymbol{X}_{i, j}: j \in S^\complement\}$ as the corresponding subsets of covariates.

Throughout the paper, we use $O_p$ and $o_p$ to denote the stochastic orders.
For a sequence of random variables $\{D_n\}$ and a deterministic sequence $\{d_n\} \in [0, \infty)$:
\begin{itemize}
    \item $D_n = o_p(d_n)$ denotes $D_n/d_n \xrightarrow{p} 0$ as $n \to \infty$. Specifically, $D_n = o_p(1)$ denotes $D_n \to 0$ in probability.
    \item $D_n = O_p(d_n)$ denotes that $\{D_n/d_n\}$ is tight: for every $\epsilon > 0$, there exists an $M > 0$ such that $\sup_n P(|D_n|/d_n > M) < \epsilon$.
\end{itemize}

\subsection{A Neyman orthogonal score} \label{s:method}

We shall make appropriate modifications to the linear adjustment approach \citep{maity2024linear} to infer $\beta^\star$. To quickly recall, the linear adjustment approach follows a two-step procedure: Firstly, it learns an estimate $\widehat \xi(\cdot)$ using the data from the majority sub-population
\begin{equation} \label{eq:predictor-majority}
   \widehat \xi = \underset{\xi \in \Xi}{\arg\min} \; \sum_{i= 1}^N T_i \ell (\xi(\boldsymbol{X}_i), Y_i) + \lambda_0 \cR_0(\xi) ,
\end{equation}
where $\cR_0(\cdot)$ is an appropriate regularization for $\xi$, and $\lambda_0 > 0$ is the regularization parameter. Then, it uses data from the minority sub-population and $\widehat \xi$ to estimate both $\beta$ and $\delta$: 
\begin{equation} \label{eq:linear-adjustment}
    \widehat \beta, \widehat \delta = \arg\min_{\beta, \delta} \; \sum_{i= 1}^N (1 - T_i) \ell (\widehat \xi(\boldsymbol{X}_i) + \beta^\top \boldsymbol{U}_i + \delta^\top \boldsymbol{V}_i, Y_i) + \lambda_1 \cR_1(\beta, \delta)\,,
\end{equation}
where $\lambda_1 > 0$ is the regularization parameter and $\cR_1$ is a regularization function for $(\beta , \delta)$.

A key advantage of this two-step procedure is its compatibility with any learning algorithm used for $\xi$. However, the estimator $\widehat \beta$ is not appropriate for our inferential goal, mainly because it may be a biased estimate of $\beta^\star$ for two reasons: (1) the finite-sample regularization applied in Eq. \eqref{eq:linear-adjustment} may shrink the estimate toward zero, and (2) the regularized estimate $\widehat\xi$ may itself be biased, which could further lead to a downstream bias in $\widehat \beta$ when calculated from Eq. \eqref{eq:linear-adjustment}. While the first source of bias can be alleviated through the debiased lasso \citep{vandegeer2014}, this correction does not generally address the second source. To handle both types of bias in a unified manner, we adopt the framework of \citet{chernozhukov2018double}, where the goal is to estimate $\beta$ through a score equation that satisfies the Neyman orthogonality condition with respect to the nuisance parameters $\xi(\cdot)$, $\delta$, and one additional nuisance parameter $\pi(\boldsymbol{X})$ with its true value $\pi^\star(\boldsymbol{X}) = \bbP(T = 1 \mid \boldsymbol{X})$. 
Following \citet[Section 2.2.1, Equation (2.15)]{chernozhukov2018double}, such a score can be constructed from the optimization in \eqref{eq:population-optimization} as
\begin{equation}\label{eq:NOS}
    S(\boldsymbol{Z}_i, \beta, \delta, \xi, \pi) = \partial _\beta \psi (\boldsymbol{Z}_i, \beta, \delta, \xi) - \boldsymbol{\Lambda} \partial_\delta \psi (\boldsymbol{Z}_i, \beta, \delta, \xi) - c(\boldsymbol{X}_i, \beta, \delta, \xi, \pi) \partial _\xi\psi (\boldsymbol{Z}_i, \beta, \delta, \xi),
    \end{equation}
where  $\boldsymbol{\Lambda} \in \bbR^{p \times (d - p)}$, $c(\boldsymbol{X}_i, \beta, \delta, \xi, \pi) \in  \bbR^p$, and $p = |S|$ is the dimension of $\beta$. Here, $\partial _\xi\psi (\boldsymbol{Z}_i, \beta, \delta, \xi)$ is a functional derivative and is formally described as: 
\begin{equation} \label{eq:functional-derivitive-def}
    \partial _\xi\psi (\boldsymbol{Z}_i, \beta, \delta, \xi) = \left[\frac{\partial }{\partial a} \left\{ T_i\ell (a, Y_i) + (1 - T_i) \ell (a +\beta^\top \boldsymbol{U}_i + \delta^\top \boldsymbol{V}_i, Y_i) \right \}\right]_{a = \xi(\boldsymbol{X}_i)}\,.
\end{equation} The score function in \eqref{eq:NOS} also contains a $\partial_\pi \psi (\boldsymbol{Z}_i, \beta, \delta, \xi)$ term, but we omit it since it equals zero. 
The $\boldsymbol{\Lambda}$ and $c(\cdot)$ are determined to satisfy the \emph{Neyman orthogonality condition}: 
\begin{equation}\label{eq:orth_condition}
\begin{aligned}
   \Ex \left [\partial _\delta S(\boldsymbol{Z},\beta^\star, \delta^\star, \xi^\star, \pi^\star)  \right] = 0, ~~ \Ex \left [\partial _\xi S(\boldsymbol{Z}, \beta^\star, \delta^\star, \xi^\star, \pi^\star)  \big \vert \boldsymbol{X} \right] = 0,\\
   \text{and} ~~ \Ex \left [\partial _\pi S(\boldsymbol{Z}, \beta^\star, \delta^\star, \xi^\star, \pi^\star)  \big \vert \boldsymbol{X} \right] = 0 .
\end{aligned}
\end{equation} The functional derivatives $\partial _\xi S(\boldsymbol{Z}_i, \beta^\star, \delta^\star, \xi^\star, \pi^\star)$ and $\partial _\pi S(\boldsymbol{Z}_i, \beta^\star, \delta^\star, \xi^\star, \pi^\star)$ are defined similarly to $\partial _\xi\psi (\boldsymbol{Z}_i, \beta, \delta, \xi)$, as in Eq. \eqref{eq:functional-derivitive-def}. The Neyman orthogonal $S$ is simplified below:
\begin{equation}\label{eq:score-function}
    \begin{aligned}
        S(\boldsymbol{Z}_i, \beta, \delta, \xi, \pi) & = \left(\boldsymbol{U}_i - \boldsymbol{\Lambda}_0 \boldsymbol{V}_i\right) \Big[(1 - T_i) \left \{\sigma \left( \xi(\boldsymbol{X}_i) + \beta^\top \boldsymbol{U}_i + \delta^\top \boldsymbol{V}_i\right) - Y_i \right\} \kappa (\boldsymbol{Z}_i, \beta, \delta, \xi, \pi)\\
        & \qquad - T_i \left \{\sigma \left( \xi(\boldsymbol{X}_i)\right) - Y_i \right\} \big(1 - \kappa (\boldsymbol{Z}_i, \beta, \delta, \xi, \pi)\big)\Big],
    \end{aligned}
\end{equation}
where 
\[
\kappa (\boldsymbol{Z}_i, \beta, \delta, \xi, \pi) = \frac{
\pi(\boldsymbol{X}_i)\,\sigma'\!\big(\xi(\boldsymbol{X}_i)\big)
}{
\pi(\boldsymbol{X}_i)\,\sigma'\!\big(\xi(\boldsymbol{X}_i)\big)
+
\{1 - \pi(\boldsymbol{X}_i)\}\,
\sigma'\!\big(\xi(\boldsymbol{X}_i) + \beta^\top \boldsymbol{U}_i + \delta^{\top} \boldsymbol{V}_i\big)
},
\]
and $\boldsymbol{\Lambda}_0$ is the minimizer described below:
\[
\boldsymbol{\Lambda}_0 = \arg\min_{\boldsymbol{\Lambda}} \Ex\left[ (1 - T)\kappa (\boldsymbol{Z}, \beta, \delta, \xi, \pi)\,
\sigma'\!\big(\xi(\boldsymbol{X}) + \beta^\top \boldsymbol{U} + \delta^{\top} \boldsymbol{V}\big) \| \boldsymbol{U} - \boldsymbol{\Lambda} \boldsymbol{V} \|^2_2 \right].
\] In Lemma S3.1,
we establish that the score function satisfies the Neyman orthogonality condition, as described in \eqref{eq:orth_condition}.
While $\boldsymbol{\Lambda}_0$ is a population quantity and cannot be calculated from the data, we shall replace it with an estimate $\widehat{ \boldsymbol{\Lambda}}$ calculated as follows: 
\begin{equation}
\begin{aligned}
    \widehat{ \boldsymbol{\Lambda}} & = \arg\min_{\boldsymbol{\Lambda}} \sum_{i=1}^N 
w_i\,\| \boldsymbol{U}_i - \boldsymbol{\Lambda} \boldsymbol{V}_i \|^2 + \lambda_2 \cR_2(\boldsymbol{\Lambda}),
\end{aligned}
\end{equation} 
where \[
w_i =
(1 - T_i)\kappa (\boldsymbol{Z}_i, \beta, \delta, \xi, \pi)\,
\sigma'\!\big(\xi(\boldsymbol{X}_i) + \beta^\top \boldsymbol{U}_i + \delta^{\top} \boldsymbol{V}_i\big).
\]
$\cR_2(\boldsymbol{\Lambda})$ is an appropriate regularization term and can be ignored for low dimensional $\boldsymbol{\Lambda}$.

With the score function defined as Eq. \eqref{eq:score-function}, we calculate a debiased estimate $\widehat \beta_{\debias}$ for $\beta^\star$ as the solution to the following score equation: 
\begin{equation}\label{eq:betahat}
    \sum_{i = 1}^N S\left(\!\boldsymbol{Z}_i, \beta, \widehat \delta, \widehat \xi, \widehat \pi\right) = 0\,,
\end{equation} where $\widehat \xi$ is calculated from  Eq \eqref{eq:predictor-majority}, $\widehat \delta$ is calculated from Eq \eqref{eq:linear-adjustment}, and $\widehat \pi$ is an estimated predictive model for $\pi^\star(\boldsymbol{X}_i) = \bbP(T_i = 1\mid \boldsymbol{X}_i)$ within a model class $\Pi$, calculated as
\[
\widehat \pi = \arg\min_{\pi \in \Pi}\;  \frac1 N\sum_{i = 1}^N \ell_T(\pi(\boldsymbol{X}_i), T_i) + \lambda_3 \cR_3(\pi)\,. 
\] 
The score equation in \eqref{eq:betahat} is an implicit equation and does not have  a closed form solution. In Algorithm \ref{ch3alg:NO}, we provide an iterative approach to approximate the solution $\widehat \beta_{\debias}$.

\subsection{Statistical inference}\label{s:inference}

In this subsection, we establish an asymptotic distribution for the debiased estimator $\widehat \beta_{\debias}$ proposed in Eq. \eqref{eq:betahat}, which is a  key component for drawing inference on $\beta^\star$. To establish the asymptotic distribution, we make the following two assumptions:

\begin{assumption}\label{Assumption:Xconstraint} 
Given an $ M  > 0 $ the covariates satisfy the uniform bound $\|\boldsymbol{X}\|_{\infty} \leq M$. 
\end{assumption}
The boundedness assumption is a sufficient condition for establishing the asymptotic distribution in Theorem \ref{thm:asymnormal} and is primarily considered to make  our technical calculations significantly easier. However, we believe that the result holds more generally; in our simulations in Section \ref{s:simulation}, we validate the asymptotic distribution of $\widehat \beta_{\debias}$ by letting $\boldsymbol{X}$ be unbounded.

\begin{assumption}\label{Assumption:constraints} Estimates of the nuisance parameters satisfy the following conditions:
     $\sqrt{N}\|\widehat\delta - \delta^{\star}\|_2^2 \to 0$, $\sqrt{N}\| \widehat\xi(\cdot) - \xi^{\star}(\cdot)\|^2_{2,\bbP_{\boldsymbol{X}}} \to 0$, and $\sqrt{N}\| \widehat\pi(\cdot) - \pi^{\star}(\cdot)\|^2_{2,\bbP_{\boldsymbol{X}}} \to 0$ in probability. 
\end{assumption}
Similar assumptions are common in the semi-parametric inference literature \citep{robins1995semiparametric,chernozhukov2018double}, which allows estimates of the nuisance  parameters to converge at a rate slower than $\sqrt{N}$. In our case, if $\delta^\star$ is a high-dimensional parameter with sparsity $s$, then an $\ell_1$-regularized estimate $\widehat \delta$ satisfies $\|\widehat \delta - \delta^\star \|_2 = \cO_p(\sqrt{s \log p/ N})$ \citep{bickel2009simultaneous,vandegeer2014}. Consequently, $\sqrt{N}\|\widehat\delta - \delta^{\star}\|_2^2 = o_p(1)$ holds whenever $s\log p / \sqrt{N} \to 0$. Furthermore, under suitable conditions, one may argue that the nearest neighbor \citep{devroye1994strong} and kernel smoothing \citep{wand1994kernel} based non-parametric estimates of $\pi^\star$ and $\xi^\star$ satisfy the above condition. In our experiments in Section \ref{s:application}, the $\xi^\star$ is estimated using XGBoost, though we note that a comprehensive theoretical study on XGBoost is limited in the literature.  While we are unable to verify the assumption theoretically for XGBoost, in our simulation (Section \ref{s:simulation}), we provide an ablation study on the effect of model misspecification.

Additionally, we require some regularity assumptions, which we present in Assumption \ref{Assumption:psi}. Under these assumptions, we formalize the asymptotic distribution of $\widehat \beta_{\debias}$ below: 
\begin{theorem}\label{thm:asymnormal}
Consider the debiased estimator $\widehat \beta_{\debias}$ defined in Eq. \eqref{eq:betahat}. Under the Assumptions \ref{Assumption:Xconstraint}, \ref{Assumption:constraints},
\ref{Assumption:psi}, and Assumption~S2 in Online Resource~1, and that 
\[
\begin{aligned}
  \boldsymbol{A} & := \Ex\left[\partial_{\beta} S(\boldsymbol{Z}, \beta^\star, \delta^\star, \xi^\star, \pi^\star)\right] \\
  & =  \Ex\left[(1-T) \sigma'(\xi^\star(\boldsymbol{X}) + {\beta^\star}^\top  \boldsymbol{U} +{\delta^\star}^\top \boldsymbol{V})  \kappa (\boldsymbol{Z}, \beta^\star, \delta^\star, \xi^\star, \pi^\star)(\boldsymbol{U}- \boldsymbol{\Lambda}_0 \boldsymbol{V} )(\boldsymbol{U}- \boldsymbol{\Lambda}_0 \boldsymbol{V} )^\top \right]
\end{aligned}
\] is invertible,  
the following holds: 
\[
\sqrt{N} \left(\widehat \beta_{\debias} - \beta^\star\right) \to \bN\left( \boldsymbol{0}, \bV \right) ~~ \text{in distribution}\,,
\] where, for $\boldsymbol\Omega  := \var \left[S(\boldsymbol{Z}, \beta^\star, \delta^\star, \xi^\star, \pi^\star)\right]$, $\bV = \boldsymbol{A}^{-1} \boldsymbol\Omega \boldsymbol{A}^{-T}$. 

\end{theorem}


For a smaller proportion of the minority subgroup, the asymptotic variance of  $\hat\beta_{debias}$ is higher. In Corollary \ref{coro:group_imbalance} in the Appendix, we formalize this property and establish that the effective sample size for inference is the same as the sample size for the minority group.

Having established the asymptotic normality for $\widehat \beta_{\debias}$, we require an estimate for $\bV$ as our final tool for inference. The $\Omega$ can be estimated as 
\[
    \widehat{\boldsymbol\Omega} = \textstyle \frac1{N-1} \sum_{i=1}^N \left \{ S\left(\!\boldsymbol{Z}_i, \widehat \beta_{\debias}, \widehat \delta, \widehat \xi, \widehat \pi\right) - \widehat \mu \right\} \left \{ S\left(\!\boldsymbol{Z}_i, \widehat \beta_{\debias}, \widehat \delta, \widehat \xi, \widehat \pi\right) - \widehat \mu \right\}^\top,
\] 
where
\(
\widehat \mu = \textstyle \frac1 N \sum_{i = 1}^N S\left(\!\boldsymbol{Z}_i, \widehat \beta_{\debias}, \widehat \delta, \widehat \xi, \widehat \pi\right).
\)
Moreover,
using its equality in Theorem \ref{thm:asymnormal}, we estimate $\boldsymbol{A}$ as
\[
{\widehat \bA} = \frac1N \sum_{i = 1}^N (1-T_i) 
\kappa_i 
\sigma'(\widehat\xi(\boldsymbol{X}_i) + \widehat\beta_{\debias}^\top  \boldsymbol{U}_i +\widehat \delta^\top \boldsymbol{V}_i)  \left(\boldsymbol{U}_i- \widehat {\boldsymbol{\Lambda}}_0 \boldsymbol{V}_i \right)\left(\boldsymbol{U}_i- \widehat {\boldsymbol{\Lambda}}_0 \boldsymbol{V}_i \right)^\top\,,
\] where $\kappa_i:=  \kappa(\boldsymbol{Z}_i, \widehat\beta_{\debias}, \hat\delta, \hat\xi, \hat\pi)$. Finally, we estimate the asymptotic variance as 
\begin{equation}
    \widehat \bV = {\widehat \bA}^{-1} \boldsymbol{\widehat \Omega} {\widehat \bA}^{-T}\,. 
\end{equation} In the following lemma, we establish consistency for the variance estimate. 
\begin{lemma} \label{lemma:asymvar} 
Under the conditions assumed in Theorem \ref{thm:asymnormal}, $\widehat \bV \to \bV$ in probability. 
\end{lemma}

\noindent\textbf{Confidence interval.} The asymptotic distribution in Theorem \ref{thm:asymnormal} and the consistency of the variance estimate in Lemma \ref{lemma:asymvar} can be utilized to construct an asymptotic confidence interval for $\beta^\star$. For example, a two-sided confidence interval for $\beta^\star_j$ with an asymptotic coverage probability of $1 - \alpha$ is
\begin{equation}\label{eq:CI}
    \beta^\star_j \in \left[\big[\widehat \beta_{\debias}\big]_j - z_{1 - \frac{\alpha}{2}} \frac{\widehat \bV^{\frac12}_{jj}}{\sqrt{N}} , \big[\widehat \beta_{\debias}\big]_j + z_{1 - \frac{\alpha}{2}} \frac{\widehat \bV^{\frac12}_{jj}}{\sqrt{N}}  \right]
\end{equation} where $z_{1 - \nicefrac{\alpha}{2}}$ is the $(1 - \nicefrac{\alpha}{2})$-th quantile of $\bN(0, 1)$.

\begin{remark}[Semi-parametric efficiency]
    If the working model in Eq. \eqref{eq:working-model} is true data generating model, under the conditions in Theorem \ref{thm:asymnormal}, the orthogonal score in \eqref{eq:score-function} coincides with the efficient score for the corresponding semi-parametric M-estimation problem (\cite{chernozhukov2018double}, Section 2.2.3), and our estimator $\hat\beta_{debias}$ is semi-parametrically efficient: its asymptotic variance is the smallest among all regular asymptotically linear estimators \citep{tsiatis2006semiparametric}, and the confidence interval is asymptotically the shortest interval with correct coverage (\cite{chernozhukov2018double}, Corollary 3.2).
\end{remark}

\section{Simulation}\label{s:simulation}

We conduct simulation studies to evaluate the finite-sample performance of the proposed debiased estimator. Specifically, we (1) compare it with the linear adjustment method (Algorithm~\ref{ch3alg:LA}) \citep{maity2024linear}, with and without $\ell_1$ regularization in the second step, under several baseline learners for $\xi(\cdot)$ in Eq.~\eqref{eq:predictor-majority}; (2) assess its robustness when the baseline model is misspecified; (3) examine how performance varies with the degree of group imbalance; and (4) empirically verify its asymptotic normality.

\subsection{Simulation setting}
We consider a total sample size of $N = 30{,}000$, which is of the same order as the eICU dataset, and three minority group proportions $p_0 = \bbP(T=0) \in \{0.1, 0.2, 0.4\}$, giving minority sample sizes $n_0 = N p_0 \in \{3000, 6000, 12000\}$. These proportions reflect typical group compositions in medical datasets. The covariates $\boldsymbol{X} = (\boldsymbol{X}_1, \dots, \boldsymbol{X}_d)$, with $d = 20$, include both continuous and discrete variables, as is common in medical data. Specifically, $\boldsymbol{X}_3 \sim \text{Bernoulli}(0.5)$, $\boldsymbol{X}_4 \sim \text{Bernoulli}(0.4)$, $\boldsymbol{X}_5 \sim \text{Bernoulli}(0.3)$, and $\boldsymbol{X}_{11} \sim \text{Bernoulli}(0.5)$ are binary; $\boldsymbol{X}_{13}$ is categorical, taking values in $\{0, 1, 2\}$ with equal probability; and the remaining covariates are independent standard normal. Since $\boldsymbol X$ is unbounded, this setup intentionally violates Assumption~\ref{Assumption:Xconstraint}, which allows us to assess robustness beyond the bounded support condition.

The outcomes are generated from a logistic regression model 
\begin{equation}\label{eq:simulation}
    \begin{aligned}
    &\mathrm{logit}\big\{\bbP(Y=1\mid \boldsymbol{X}, T = 1)\big\} = \xi^\star(\boldsymbol X), \\
    &\mathrm{logit}\big\{\bbP(Y=1\mid \boldsymbol{X}, T = 0)\big\} = \xi^\star(\boldsymbol X) + \beta \boldsymbol{X}_1 + \delta \boldsymbol{X}_2,
    \end{aligned}
\end{equation}
with $\beta^\star = \delta^\star = 0.5$; our target is $\beta$. We consider two baseline models:
\begin{equation}\label{eq:sim-baseline}
    \begin{aligned}
    &\text{Scenario 1 (linear):} && \xi^\star(\boldsymbol X) = \mathrm{logit}(0.1) - \frac{3}{10}\sum_{j=1}^{10}\boldsymbol{X}_j, \\
    &\text{Scenario 2 (nonlinear):} && \xi^\star(\boldsymbol X) = \mathrm{logit}(0.1) - \frac{3}{10}\sum_{j=1}^{10}\boldsymbol{X}_j + \frac{1}{4}\boldsymbol{X}_6\boldsymbol{X}_7 + \frac{1}{4}\big(\boldsymbol{X}_8^2 - 1\big).
    \end{aligned}
\end{equation}
Each configuration is replicated 1000 times.

For Scenario 1, we fit the baseline $\xi(\cdot)$ on the majority group with three learners: a penalized GLM, XGBoost, and a single-hidden-layer neural network. For Scenario 2, we fit only the penalized GLM, which is linear in $\boldsymbol X$ and therefore misspecified. Given $\hat\xi(\cdot)$, we compare three estimators of $\beta$:
(i) the linear adjustment estimator $\hat\beta_{\mathrm{LA}}$, obtained from Eq.~\eqref{eq:linear-adjustment} with an $\ell_1$ penalty on $(\beta, \delta)$;
(ii) the unpenalized linear adjustment estimator $\hat\beta_{\mathrm{unpen}}$, obtained from Eq.~\eqref{eq:linear-adjustment} with $\lambda_1 = 0$, i.e., the low-dimensional branch of Algorithm~\ref{ch3alg:LA}; and 
(iii) the proposed debiased estimator $\hat\beta_{\debias}$ (Algorithm~\ref{ch3alg:NO}).
The estimator $\hat\beta_{\mathrm{unpen}}$ removes the regularization bias but still treats $\hat\xi(\cdot)$ as known. Comparing the three estimators therefore separates the two sources of bias discussed in Section~\ref{s:method}: regularization and the estimation error in $\hat\xi(\cdot)$.

We report the absolute bias, the standard deviation, and the empirical coverage of 95\% confidence intervals. Confidence intervals are constructed for $\hat\beta_{\debias}$ using Eq.~\eqref{eq:CI} and for $\hat\beta_{\mathrm{unpen}}$ using Wald intervals from the second-step logistic regression with $\hat\xi(\cdot)$ treated as a fixed offset. No interval is reported for $\hat\beta_{\mathrm{LA}}$, because the lasso estimator does not admit standard inference.


\subsection{Scenario 1 result: linear baseline}\label{s:sim-linear}

\begin{figure}[H]
    \centering
    \includegraphics[width=\linewidth]{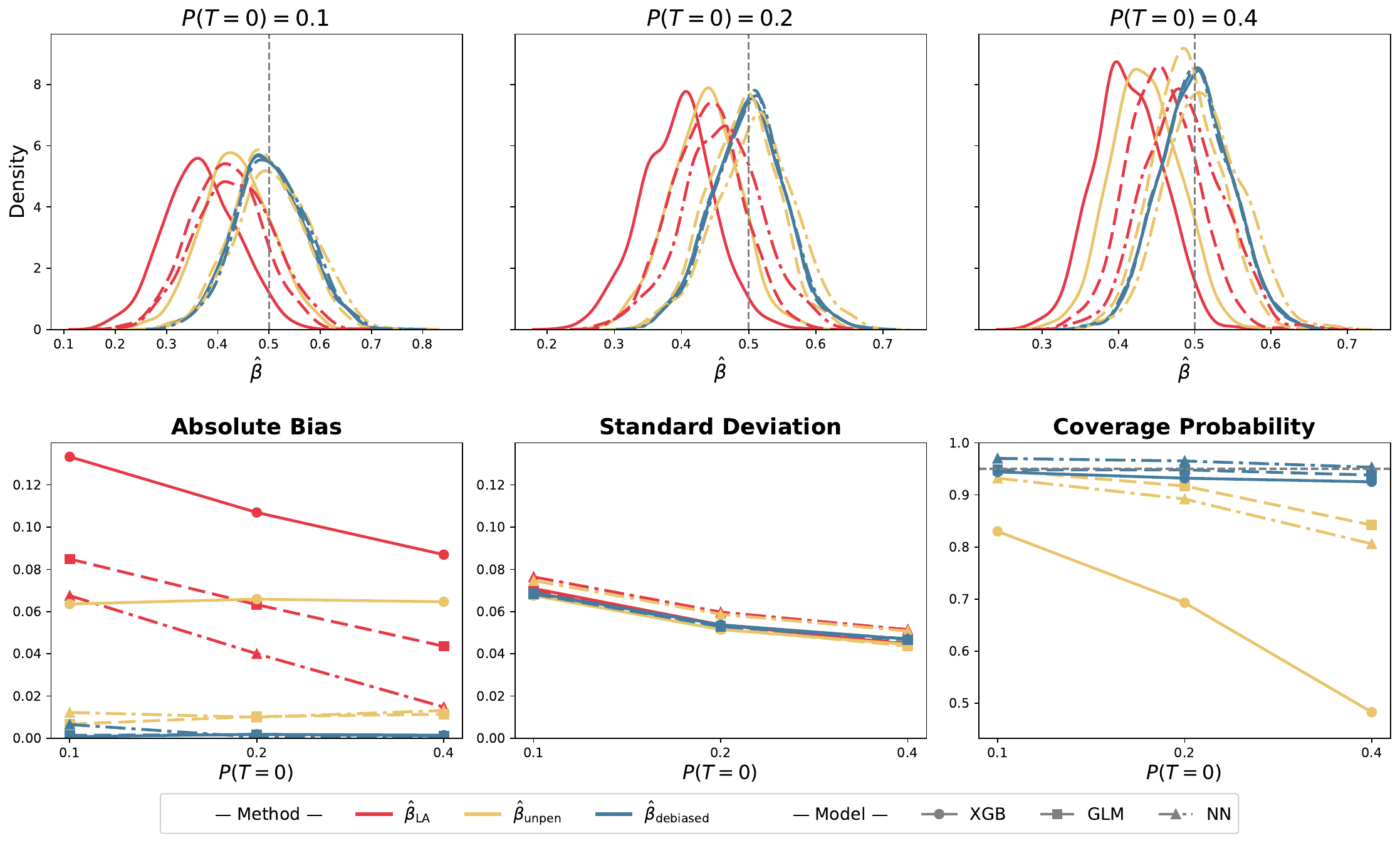}
    \caption{Scenario 1 (linear baseline): results for $\hat\beta_{\mathrm{LA}}$ (red), $\hat\beta_{\mathrm{unpen}}$ (yellow), and $\hat\beta_{\debias}$ (blue) across 1000 replicates. Top row: smoothed densities of the estimates for $\bbP(T=0) = 0.1$, $0.2$, and $0.4$; the vertical dashed line marks the true value $\beta^\star = 0.5$. Bottom row: absolute bias, standard deviation, and 95\% confidence interval coverage (for $\hat\beta_{\mathrm{unpen}}$ and $\hat\beta_{\debias}$ only) as functions of $\bbP(T=0)$; the horizontal dashed line marks the nominal level 0.95. Line types and markers indicate the baseline learner: XGBoost (XGB, solid, circles), penalized GLM (GLM, dashed, squares), and single-hidden-layer neural network (NN, dash-dot, triangles)}
    \label{fig:sim_linear}
\end{figure}

Figure~\ref{fig:sim_linear} shows the sampling distributions of the three estimators (panel~a) and summarizes their bias, standard deviation, and coverage (panel~b).
The debiased estimator $\hat\beta_{\mathrm{debiased}}$ is centered at the true value, with near-zero bias for all baseline learners and all levels of imbalance. The lasso estimator $\hat\beta_{\mathrm{LA}}$ is shifted toward zero and has the largest bias. This bias decreases as $\bbP(T=0)$ increases, consistent with weaker $\ell_1$ shrinkage for larger minority samples, but it remains non-negligible.
Removing the $\ell_1$ penalty moves the sampling distribution of $\hat\beta_{\mathrm{unpen}}$ toward the truth, but a visible shift remains when the baseline model is fitted with XGBoost -- This residual bias comes from the estimation error in $\hat\xi(\cdot)$, which is learned from the majority group and does not diminish as the minority sample grows. Thus removing the penalty addresses only the regularization bias, whereas the debiasing correction addresses both sources.
In addition, the bias of the debiased estimator remains substantially smaller than its standard deviation across all imbalance levels, which is favorable for valid inference. 
The 95\% confidence intervals constructed from the debiased estimator achieve empirical coverage close to the nominal level across all imbalance settings and baseline models. The intervals based on $\hat\beta_{\mathrm{unpen}}$ undercover, and their coverage deteriorates as $\bbP(T=0)$ increases. 
In the Appendix~\ref{append:qq}, the QQ-plots (Figure \ref{fig:sim_qqplot}) demonstrate the approximate normality of the debiased estimator from this simulation study.

\subsection{Scenario 2 result: misspecified non-linear baseline}\label{s:sim-misspecified}

\begin{figure}[H]
    \centering
    \includegraphics[width=\linewidth]{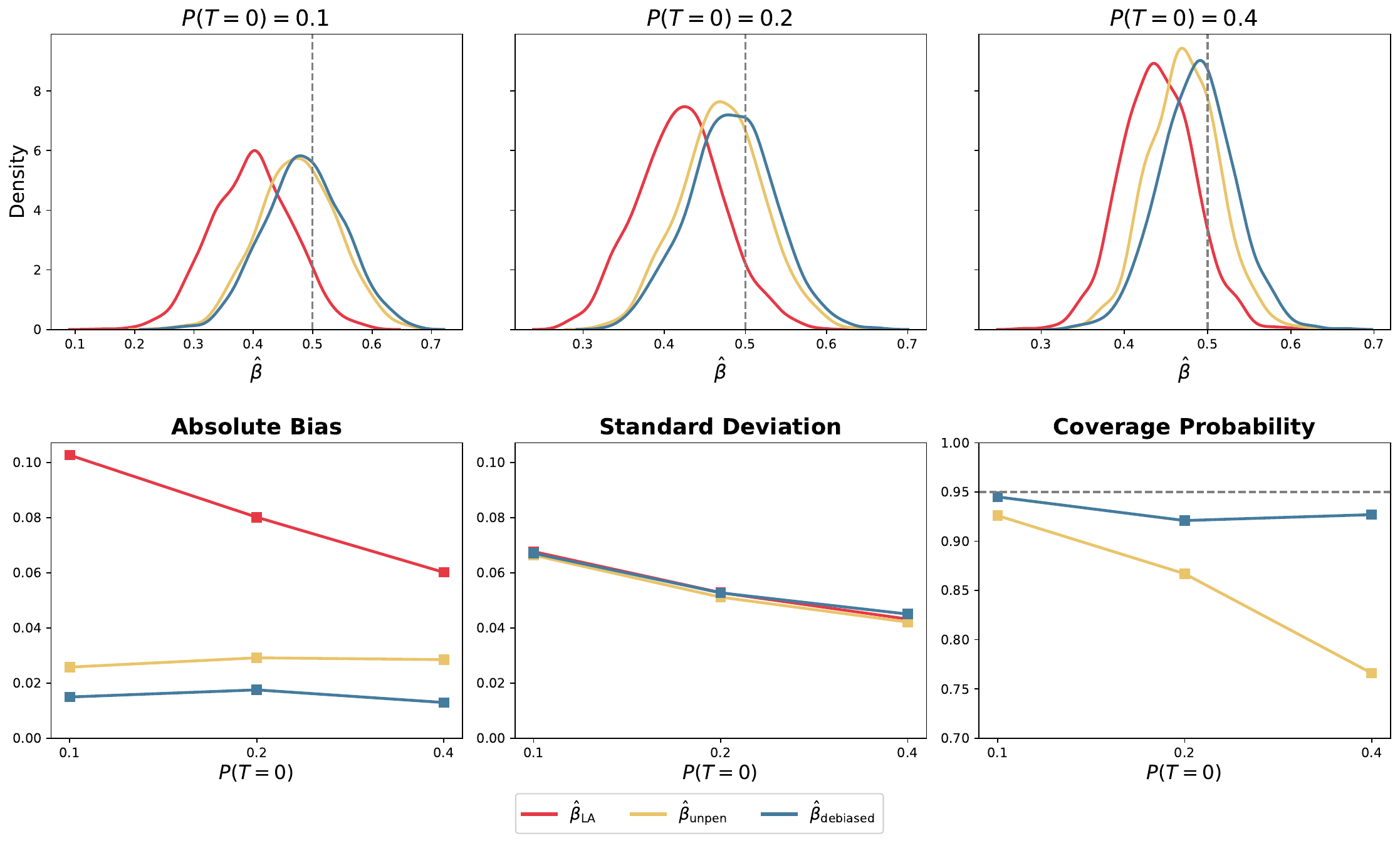}
    \caption{Scenario 2 (misspecified GLM baseline): results for $\hat\beta_{\mathrm{LA}}$ (red), $\hat\beta_{\mathrm{unpen}}$ (yellow), and $\hat\beta_{\debias}$ (blue) across 1000 replicates. The layout is the same as in Figure~\ref{fig:sim_linear}}
    \label{fig:sim_nonlinear}
\end{figure}

In Scenario 1, the GLM baseline is correctly specified. Scenario 2 examines what happens when it is not: the GLM cannot capture the interaction and quadratic terms in $\xi^\star(\cdot)$.  Figure~\ref{fig:sim_nonlinear} shows the results.
The debiased estimator has the smallest bias in every setting, remaining well below the standard deviation, and its coverage stays between 0.92 and 0.95. Thus, inference based on $\hat\beta_{\debias}$ remains approximately valid under baseline misspecification (see Figure~\ref{fig:sim_qqplot_nonlinear} for the corresponding QQ plots). 
However, the lasso estimator $\hat\beta_{\mathrm{LA}}$ is again shifted toward zero. $\hat\beta_{\mathrm{unpen}}$ reduces the bias by removing the penalty, but this bias is still not negligible. In addition, the coverage of $\hat\beta_{\mathrm{unpen}}$ falls from 0.93 to 0.77. 

\section{Data application}\label{s:application}
In the ICU clinical setting, many interventions follow standardized protocols \citep{siner2015protocol} that are applied broadly rather than tailored to individual ethnicity, so subgroup differences in outcomes may go unnoticed or unaddressed \citep{mcgowan2022racial, kotfis2024equity}. To investigate this issue, we use the publicly available eICU database \citep{pollard2018eicu} to examine heterogeneity in in-hospital mortality between Caucasian patients and three under-represented ethnic groups: Native American, Asian, and Hispanic (Table \ref{tab:mortal_all}). We use Caucasians as the baseline group because they are well represented, with far more observations than the other groups. The three under-represented groups have much smaller sample sizes, which makes direct estimation more difficult and may amplify estimated disparities. Accurately characterizing these differences can give a more reliable picture of mortality risk across populations and support more targeted clinical strategies. The analysis includes $d = 67$ predictors, which is large relative to the number of deaths in each minority group (42 to 68). We compare the proposed debiased estimator with the two versions of the linear adjustment approach \citep{maity2024linear} used in Section~\ref{s:simulation}: with an $\ell_1$ penalty ($\hat\beta_{\mathrm{LA}}$) and without a penalty ($\hat\beta_{\mathrm{unpen}}$). 

\begin{table}[htbp]
\centering
\caption{In-hospital mortality by ethnicity in the eICU data. $\widehat{p}_{\text{minority}} - \widehat{p}_{\text{Caucasian}}$ denotes the estimated difference in mortality prevalence between each minority subgroup and the Caucasian reference subgroup}
\begin{tabular}{lcccc}
\toprule
Status & Caucasian & Asian & Hispanic & Native American \\
\midrule
Alive   & 33427 & 473 & 875 & 425 \\
Expired & 3616  & 49  & 68  & 42  \\
\midrule
Total   & 37043 & 522 & 943 & 467 \\
Prevalence (\%) & 9.77 & 9.39 & 7.21 & 8.99 \\
\midrule
$\widehat p_{minority} - \widehat p_{Caucasian}$ & - & -0.0037 & -0.0255 & -0.0077 \\
P-value$^{a}$ & - & 0.3854 & 0.0014 & 0.2822 \\
\bottomrule
\end{tabular}

\vspace{2pt}
\begin{minipage}{\linewidth}\footnotesize
$^{a}$P-value is calculated for the alternative $H_1: p_{\mathrm{minority}} < p_{\mathrm{Caucasian}}$.
\end{minipage}
\label{tab:mortal_all}
\end{table}

\subsection{Result}
Table~\ref{tab:Allethnicity} reports estimates of coefficient heterogeneity between Caucasians and the under-represented ethnic groups for specific admission diagnoses. The two linear adjustment estimators fail for different reasons. The lasso estimates $\hat\beta_{\mathrm{LA}}$ (Eq.~\eqref{eq:linear-adjustment}) are all shrunk exactly to zero, and removing the penalty does not solve the problem. For Overdose among Native Americans and Cardiac Arrest among Asians, the unpenalized estimates $\hat\beta_{\mathrm{unpen}}$ diverge (about $-130$ and $-78$), with standard errors of order $10^5$. This indicates that the unpenalized maximum likelihood estimate effectively does not exist, because there are too few deaths within these diagnosis subgroups of the minority groups relative to the large number of covariates ($d = 67$). For diabetic ketoacidosis (DKA) among Hispanics, $\hat\beta_{\mathrm{unpen}}$ is finite, but its standard error ($2.82$) is too large to detect any effect ($p = 0.28$). In contrast, the debiased estimator $\hat\beta_{\debias}$ yields stable estimates and identifies significant positive heterogeneity in all three cases. For example, $\hat\beta_{\debias} = 1.22$ for Native American patients admitted with Overdose ($p = 0.02$): after adjusting for other covariates, these patients had higher log-odds of in-hospital mortality than comparable Caucasian patients. Similarly, the debiased estimator indicates poorer Cardiac Arrest outcomes among Asian patients and poorer DKA-related outcomes among Hispanic patients, both consistent with existing literature \citep{sherrod2025survival,gupta2023comparison,ghobrial2016ethnic,nyenwe2007admissions,everett2019association,bergmann2022association,fernandez2011language}.

\begin{table}[htbp]
\centering
\caption{Inference results on coefficient heterogeneity between Caucasians and under-represented ethnic groups. 
Est (SE): point estimate with standard error; $p$: one-sided p-value for $H_0:\beta=0$ vs.\ $H_1:\beta>0$.}
\label{tab:Allethnicity}
\small
\setlength{\tabcolsep}{4pt}
\begin{tabular}{@{}l c cc ccc@{}}
\toprule
& $\hat\beta_{\mathrm{LA}}$
  & \multicolumn{2}{c}{$\hat\beta_{\mathrm{unpen}}$}
  & \multicolumn{3}{c}{$\hat\beta_{\debias}$} \\
\cmidrule(lr){2-2}\cmidrule(lr){3-4}\cmidrule(l){5-7}
Group / Diagnosis & Est & Est (SE) & $p$ & Est (SE) & 95\% CI & $p$ \\
\midrule
\multicolumn{7}{@{}l}{\textit{Native American}} \\
\quad Overdose       & 0.00 & \multicolumn{2}{c}{diverged$^{a}$} & 1.22 (0.59) & (0.06, 2.38) & 0.02 \\
\multicolumn{7}{@{}l}{\textit{Asian}} \\
\quad Cardiac Arrest & 0.00 & \multicolumn{2}{c}{diverged$^{a}$} & 1.00 (0.48) & (0.05, 1.95) & 0.02 \\
\multicolumn{7}{@{}l}{\textit{Hispanic}} \\
\quad DKA            & 0.00 & 1.63 (2.82) & 0.28 & 2.36 (0.75) & (0.90, 3.82) & $<$0.01 \\
\bottomrule
\end{tabular}

\vspace{2pt}
\begin{minipage}{\linewidth}\footnotesize
$^{a}$The unpenalized estimate does not converge: $\hat\beta_{\mathrm{unpen}} = -129.59$ (SE $4.4\times10^{5}$) for Overdose and $-78.19$ (SE $1.5\times10^{5}$) for Cardiac Arrest; both one-sided $p = 0.50$. The 95\% CI of $\hat\beta_{\mathrm{unpen}}$ for DKA is $(-3.89, 7.16)$.
\end{minipage}
\end{table}


\section{Discussion}\label{s:discussion}

In this paper, we propose a debiased inference framework for detecting group-level risk heterogeneity. Our approach explicitly accommodates imbalanced group representation by using the majority group as a reference population, since this group typically benefits from well-established treatment protocols and predictive pipelines. By reliably inferring how the covariate-specific risk of minority groups differs from this majority baseline, our framework provides a robust statistical tool to inform precision medicine efforts for under-represented populations.

Central to our methodology is a Neyman orthogonal score, which makes the framework robust to estimation errors in the nuisance components. Our simulation studies separate two sources of bias in the linear adjustment approach. The $\ell_1$ penalty shrinks the estimates toward zero. Removing the penalty eliminates this regularization bias, but the estimation error in the majority-group baseline model still propagates to the estimate. This residual bias does not decrease with the minority sample size, so the confidence intervals of the unpenalized estimator undercover more severely as the minority group grows. The proposed debiased estimator corrects both sources of bias: it achieves near-zero bias and close to nominal coverage across all group imbalance levels and baseline models, with variability comparable to that of the linear adjustment estimators. When the baseline model is misspecified, the debiased estimator retains only a small residual bias and close to nominal coverage, whereas the coverage of the unpenalized estimator deteriorates further. The eICU application further highlights the practical value of our method. The lasso-based linear adjustment shrinks all estimates of interest to zero. Its unpenalized counterpart either fails to converge or yields standard errors too large to be informative, because events are sparse within diagnosis-specific minority subgroups. In contrast, our framework produces stable estimates and identifies clinically meaningful heterogeneity in diagnosis-specific mortality risk for Asian, Hispanic, and Native American patients relative to the Caucasian majority.

A key advantage of our inference framework is its flexibility in the choice of the baseline model. Any estimator of $\xi(\cdot)$ that satisfies the convergence rate conditions in Assumption \ref{Assumption:constraints} is admissible, including flexible machine learning methods such as XGBoost and neural networks. Furthermore, although our exposition focuses on binary outcomes with a logistic link function, the methodology is developed for general link ($\gamma$) and loss ($\ell$) functions. It therefore applies directly to continuous outcomes (\eg\ identity link and squared loss) and count data (\eg\ logarithmic link and Poisson deviance).

Several directions remain open for future research. First, the current framework handles binary, continuous, and count outcomes through a generalized linear model formulation. Extending it to time-to-event data (e.g., via Cox proportional hazards models) and longitudinal outcomes (e.g., via generalized estimating equations or mixed-effects models) would broaden its clinical applicability. Second, our formulation assumes that the heterogeneity takes the linear form $\alpha^\top X$. Relaxing this assumption to allow nonparametric or semiparametric specifications would enable a more flexible characterization of subgroup differences.




\section*{Data and code availability}

\paragraph{Data availability}
The data analysed in this study are available from the eICU Collaborative Research Database, version 2.0 \citep{pollard2018eicu,pollard2019eicudata}, hosted on PhysioNet \citep{goldberger2000physiobank} at \href{https://doi.org/10.13026/C2WM1R}{https://doi.org/10.13026/C2WM1R}. Access requires a credentialed PhysioNet account and a signed data use agreement; details are available at \href{https://eicu-crd.mit.edu/gettingstarted/access/}{https://eicu-crd.mit.edu/gettingstarted/access/}.


\paragraph{Code availability}
The code implementing the proposed method, the simulation studies, and the data analysis is publicly available at 
\url{https://github.com/mengqi-xu/RobustInference_RHGI}.

\bigskip 

\appendix


\section{Extra assumption and Corollary}\label{Appendix:A}

Below we state the regularity conditions on $\psi(\boldsymbol Z,\xi,\beta,\delta)$ required for establishing the asymptotic normality in Theorem \ref{thm:asymnormal} in Section \ref{s:inference}.

\begin{assumption}[Regularity conditions on $\psi$] \label{Assumption:psi}
    The loss function $\psi(\boldsymbol Z,\xi,\beta,\delta)$ satisfies the following conditions:

\begin{enumerate}
\item \textbf{Smoothness.}
$\psi(\boldsymbol Z,\xi,\beta,\delta)$ is continuously differentiable with respect to $(\xi,\beta,\delta)$ in a neighborhood of the true values $(\xi^\star,\beta^\star,\delta^\star)$.
Moreover, the partial derivatives $\partial_\beta\psi(\boldsymbol Z,\xi,\beta,\delta)$,
$\partial_\delta\psi(\boldsymbol Z,\xi,\beta,\delta)$, and
$\partial_\xi\psi(\boldsymbol Z,\xi,\beta,\delta)$
are jointly continuous in $(\xi,\beta,\delta)$.

\item \textbf{Finite moments.}
There exists $\alpha>0$ such that
\[
\Ex\!\left[
\sup_{(\xi,\beta,\delta)\in\mathcal N}
\left(
\bigl\|
\partial_\beta\psi(\boldsymbol Z,\xi,\beta,\delta)
\bigr\|^{2+\alpha}
+
\bigl\|
\partial_\delta\psi(\boldsymbol Z,\xi,\beta,\delta)
\bigr\|^{2+\alpha}
+
\bigl\|
\partial_\xi\psi(\boldsymbol Z,\xi,\beta,\delta)
\bigr\|^{2+\alpha}
\right)
\right]
<\infty,
\]
for some neighborhood $\mathcal N$ of
$(\xi^\star,\beta^\star,\delta^\star)$.

\item \textbf{Regularity of $c(\cdot)$.} $c(\cdot)$ is measurable and bounded almost surely.


\end{enumerate}
\end{assumption}



The following corollary to Theorem \ref{thm:asymnormal} formalizes the intuition that severe group imbalance --- i.e., when the minority group proportion $p_0 \to 0$ ---  inflates the asymptotic variance of $\hat\beta_{debias}$; the proof can be found in Online Resource~1. 

\begin{corollary}[Group imbalance increases variance]\label{coro:group_imbalance}
  Define $p_0 := \bbP(T = 0)$ and assume that the conditional distribution $(\boldsymbol{X}, Y | T = t)$ remains the same as we vary $p_0$.  Under the conditions of Theorem \ref{thm:asymnormal}, the $\boldsymbol{V}  p_0 $ converges to a constant as $p_0 \to 0$. Denoting the limiting constant as $\boldsymbol{V}_0:= \lim_{t_0 \to 0} \boldsymbol{V}  p_0$, it holds \[
  \sqrt{N p_0} \left(\widehat \beta_{\debias} - \beta^\star\right) \to \bN\left( \boldsymbol{0}, \bV_0 \right)~~ \text{in distribution}\,,
  \] as $p_0 \to 0$. Consequently, for small $p_0$, the effective sample size is $Np_0$. 

\end{corollary}
\section{Algorithms}

Algorithms \ref{ch3alg:NO} and \ref{ch3alg:LA} below outline the debiased inference method and linear adjustment method \citep{maity2024linear}, respectively.

\begin{algorithm}[htbp]
\caption{Debiased inference framework}\label{ch3alg:NO}
\begin{algorithmic}
\Require Observations $(Y_i, T_i, \boldsymbol{X}_i)$, $i = 1, \cdots, N$, nuisance parameter estimates $\hat\xi(\boldsymbol{X}), \hat\delta, \hat\pi(\boldsymbol{X})$, initial estimate $\beta^{(0)}$, and max iteration $R$.

\State \(
\kappa_i \gets 
\frac{
\hat\pi(\boldsymbol{X})\,\sigma'\!\big(\hat\xi(\boldsymbol{X}_i)\big)
}{
\hat\pi(\boldsymbol{X})\,\sigma'\!\big(\hat\xi(\boldsymbol{X}_i)\big)
+
\{1 - \hat\pi(\boldsymbol{X})\}\,
\sigma'\!\big(\hat\xi(\boldsymbol{X}_i) + \beta^{(0)} \boldsymbol{U}_i + \hat\delta^{\top} \boldsymbol{V}_i\big)
}.
\)
\State Set $r \gets 1$ and $\epsilon \gets 1$
\While{$r<R$ and $\epsilon > 10^{-4}$}

\(
w_i \gets
\kappa_i\,
\sigma'\!\big(\hat\xi(\boldsymbol{X}_i) + \beta^{(r-1)} \boldsymbol{U}_i + \hat\delta^{\top} \boldsymbol{V}_i\big).
\) \Comment{Observation-level weights}

\If{$\boldsymbol{V}$ is low dimensional}
    \State $\widehat{\boldsymbol{\Lambda}}
\;\gets \;
\arg\min_{\boldsymbol{\Lambda}}
\sum_{i=1}^N 
w_i\,\| \boldsymbol{U}_i - \boldsymbol{\Lambda} \boldsymbol{V}_i \|^2 , 
$ \Comment{Solve the weighted linear regression}
\ElsIf{$\boldsymbol{V}$ is high dimensional}
    \State $\widehat{\boldsymbol{\Lambda}}
\;\gets \;
\arg\min_{\boldsymbol{\Lambda}}
\sum_{i=1}^N 
w_i\,\| \boldsymbol{U}_i - \boldsymbol{\Lambda} \boldsymbol{V}_i \|^2 + \lambda\| \boldsymbol{\Lambda}\|_1
$ \Comment{Solve penalized weighted linear regression}
\EndIf

$\begin{aligned}
    S^{(r)}_i(\beta,\delta) \gets
(\boldsymbol{U}_i - \widehat{\boldsymbol{\Lambda}}\boldsymbol{V}_i)\Big[
(1 - T_i)\{\sigma(\hat\xi(\boldsymbol{X}_i)+\beta^{(r-1)} \boldsymbol{U}_i+\hat\delta^{\top} \boldsymbol{V}_i)-Y_i\}\kappa_i
\\
- T_i \{\sigma(\hat\xi(\boldsymbol{X}_i))-Y_i\}(1 - \kappa_i)
\Big].
\end{aligned}$

$\beta^{(r)} \gets \beta^{(r-1)} - \sum_{i=1}^N {S^{(r)}_i}/\sum_{i=1}^N {\partial_{\beta}S^{(r)}_i}$ 
\Comment{Update targeted estimator}
\State $r \gets r+1$, $\epsilon \gets | \beta^{(r)} - \beta^{(r-1)}|$
\Comment{Update index and convergence error}
\EndWhile
\State $\hat\beta \gets \beta^{(r)}$ \Comment{Final estimates}
\State {\bf Return:} $\hat\beta$


\end{algorithmic}
\end{algorithm}

\clearpage
\begin{algorithm}[htbp]
\caption{Linear adjustment method}\label{ch3alg:LA}
\begin{algorithmic}
\Require Observations $(Y_i, T_i, \boldsymbol{U}_i, \boldsymbol{V}_i, \boldsymbol{X}_i)$, $i = 1, \cdots, N$.
\State $\hat\xi(\boldsymbol{X}_i) \gets \arg\min_{\xi} \sum_{i=1}^N T_i\ell(\xi(\boldsymbol{X}_i), Y_i)$ \Comment{Get baseline model}
\If{$\boldsymbol{X}$ is low dimensional}
    \State $\hat\beta,\hat\delta \gets \arg\min_{\beta,\delta} \sum_{i=1}^N (1-T_i) \ell(\hat\xi(\boldsymbol{X}_i)+\beta \boldsymbol{U}_i + \delta \boldsymbol{V}_i, Y_i)$ 
\ElsIf{$\boldsymbol{X}$ is high dimensional}
    \State $\hat\beta,\hat\delta \gets \arg\min_{\beta,\delta} \sum_{i=1}^N (1-T_i) \ell(\hat\xi(\boldsymbol{X}_i)+\beta \boldsymbol{U}_i + \delta \boldsymbol{V}_i, Y_i)+ \lambda \| (\beta, \delta)\|_1$ \Comment{Final estimates}
\EndIf
\State {\bf Return:} $\hat\xi, \hat\beta, \hat\delta$
\end{algorithmic}
\end{algorithm}

\section{QQ plots of the simulation study}\label{append:qq}

\begin{figure}[H]
    \centering
    \includegraphics[width=0.9\linewidth]{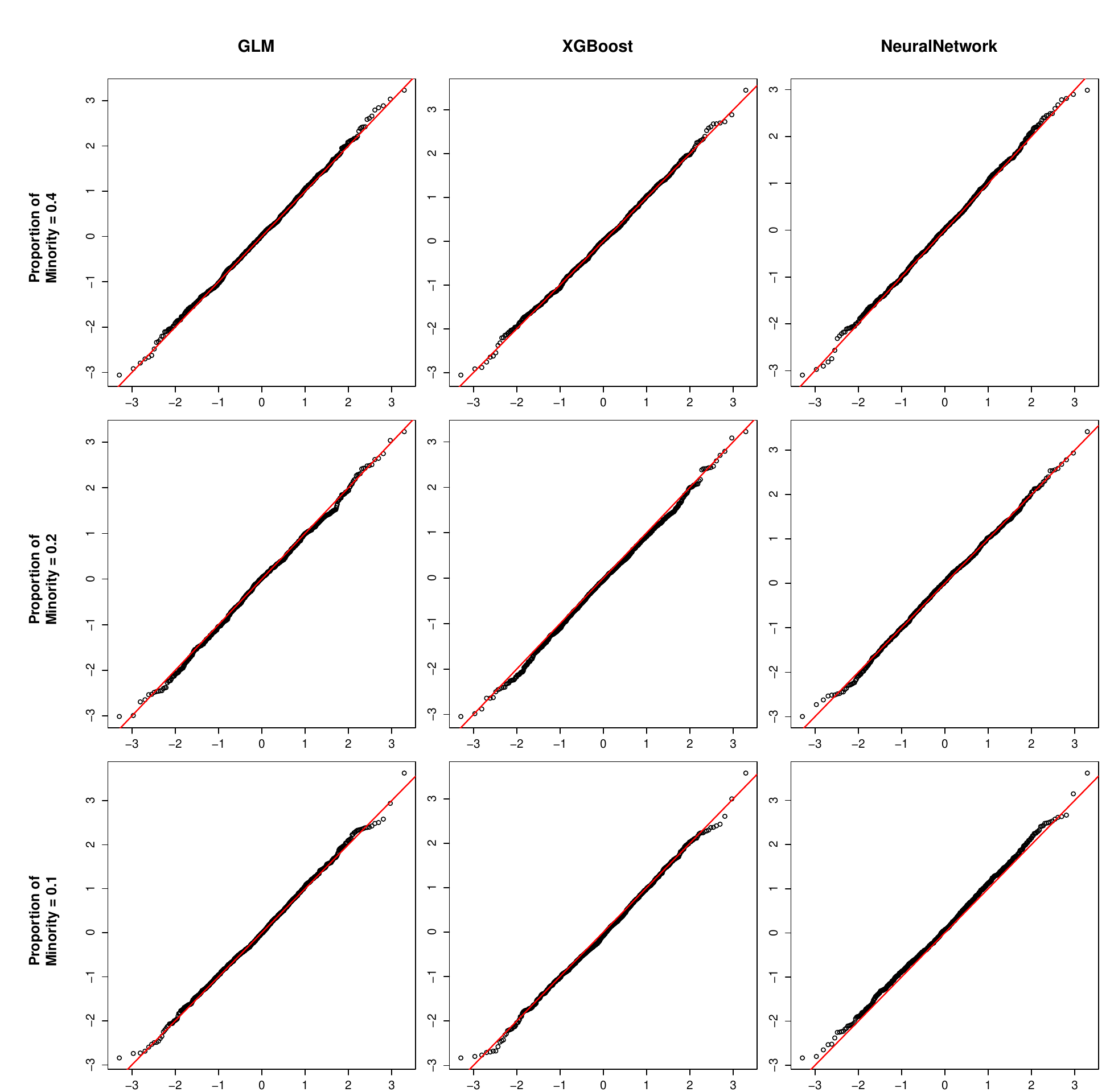}
    \newline
    \vspace{-1em}
    \caption{Normal quantile-quantile (QQ) plots of the standardized debiased estimator across 1000 simulation replicates under Scenario 1 (linear baseline). Columns correspond to three nuisance function learners: generalized linear model (GLM, left), XGBoost (middle), and neural network (right). Rows correspond to three minority group proportions $p_0 \in \{0.4, 0.2, 0.1\}$ (top to bottom). The horizontal axis shows theoretical quantiles of the standard normal distribution; the vertical axis shows the corresponding sample quantiles of the standardized estimator. The red diagonal line is the reference line of exact normality. Points align closely with the reference line across all configurations, supporting the asymptotic normality of the proposed estimator}
    \label{fig:sim_qqplot}
\end{figure}

\begin{figure}[H]
    \centering
    \includegraphics[width=\linewidth]{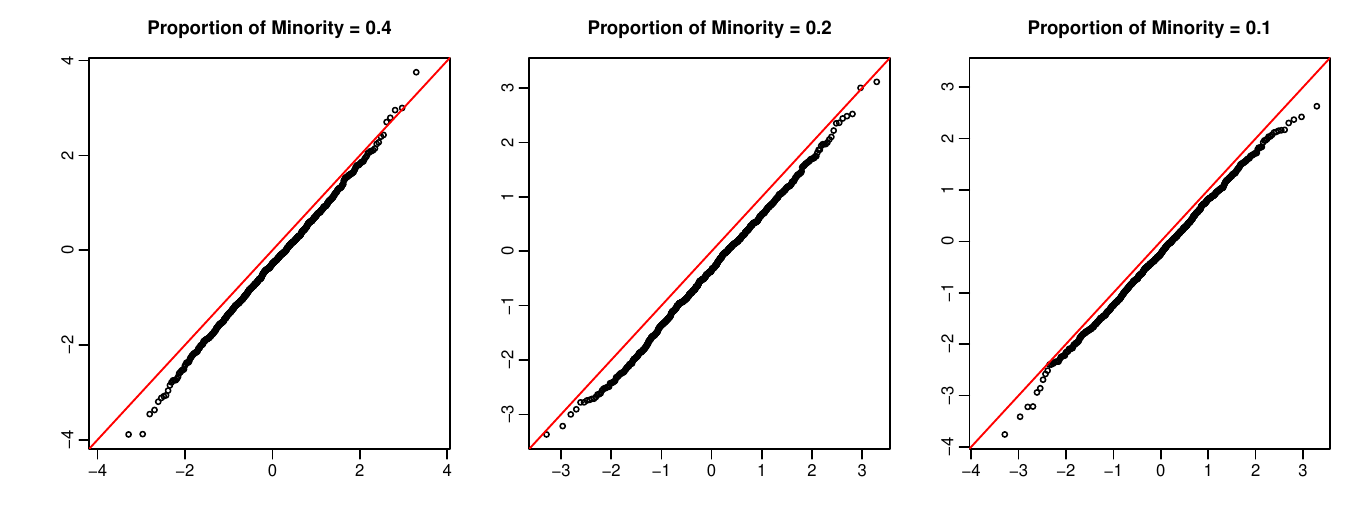}
    \caption{Normal QQ plots of the standardized debiased estimator $(\hat\beta_{\debias} - \beta^\star)/\widehat{\mathrm{SE}}$ under Scenario 2 (misspecified GLM baseline) across 1000 simulation replicates, for minority group proportions $p_0 \in \{0.4, 0.2, 0.1\}$ (left to right). The red diagonal line is the reference line of exact normality}
    \label{fig:sim_qqplot_nonlinear}
\end{figure}

Figures~\ref{fig:sim_qqplot} and~\ref{fig:sim_qqplot_nonlinear} display normal quantile-quantile (QQ) plots of the standardized debiased estimator $(\hat\beta_{\debias} - \beta^\star)/\widehat{\mathrm{SE}}$ for the two simulation scenarios in Section~\ref{s:simulation}.

Under Scenario 1, the points lie close to the reference line for every baseline learner and every minority proportion, with only minor deviations in the extreme tails. This supports the asymptotic normality established in Theorem~\ref{thm:asymnormal} and indicates that the estimated standard errors are well calibrated.

Under Scenario 2, the points are parallel to the reference line with slope close to one, so the debiased estimator remains approximately normal and its standard error correctly reflects its variability. The small downward shift corresponds to the residual bias induced by the misspecified baseline, which is small enough that the confidence intervals retain near-nominal coverage.

\clearpage
\bibliographystyle{apalike}
\bibliography{ref}

\end{document}